\documentclass{cimento}
\usepackage{graphicx}

\title{Early GRB afterglow from a reverse shock as a tracer of the prompt gamma-ray light curve}

\author{Ehud Nakar\from{ins:caltech},
Tsvi Piran\from{ins:caltech}\from{ins:huji}}

 \instlist{\inst{ins:caltech}Theoretical astrophysics, Caltech, Pasadena, 91125, USA
  \inst{ins:huji} Racah Institute for Physics, The Hebrew
University, Jerusalem 91904, Israel}

\begin{document}

\maketitle

\begin{abstract}
We discuss the optical and radio early afterglow emission of the
reverse shock that crosses a baryonic ejecta as it interacts with
the external interstellar medium (ISM). We show that the peak of the
optical flash divides the light curve of the reverse shock into two
distinctive phases. The emission after the peak depends weakly on
the initial conditions of the ejecta and therefore it can be used as
an identifiable signature of a reverse shock emission. On the other
hand, the emission before the optical peak is highly sensitive to
the initial conditions and therefore can be used to investigate the
initial hydrodynamic profile of the ejecta. In particular, if the
prompt $\gamma$-ray emission results from internal shocks, the early
reverse shock emission should resemble a smoothed version of the
prompt $\gamma$-ray light curve.
\end{abstract}

\section{Introduction}

According to the internal-external shocks model the prompt gamma-ray
burst (GRB) is produced by internal shocks within a relativistic
flow while the afterglow is produced by external shocks between this
flow and the surrounding matter. The early afterglow appears during
the transition from the prompt $\gamma$-ray emission to the
afterglow. During this transition the relativistic flow, ejected  by
the source, interacts directly with the circum burst medium. This
interaction can be used to pin down the nature of the relativistic
flow (baryonic or Poynting flux). In a baryonic flow the reverse
shock (RS) that propagates into the ejecta produces both optical and
radio emission. A Poynting flux flow is expected to produce only the
higher energy forward shock emission. While other sources of early
optical and radio emission may exist also in Poynting Flux, the RS
emission has a very robust optical and radio observable signatures
that are very unlikely to be imitated by other phenomena
(\cite{NP04}).

\begin{figure}
\includegraphics{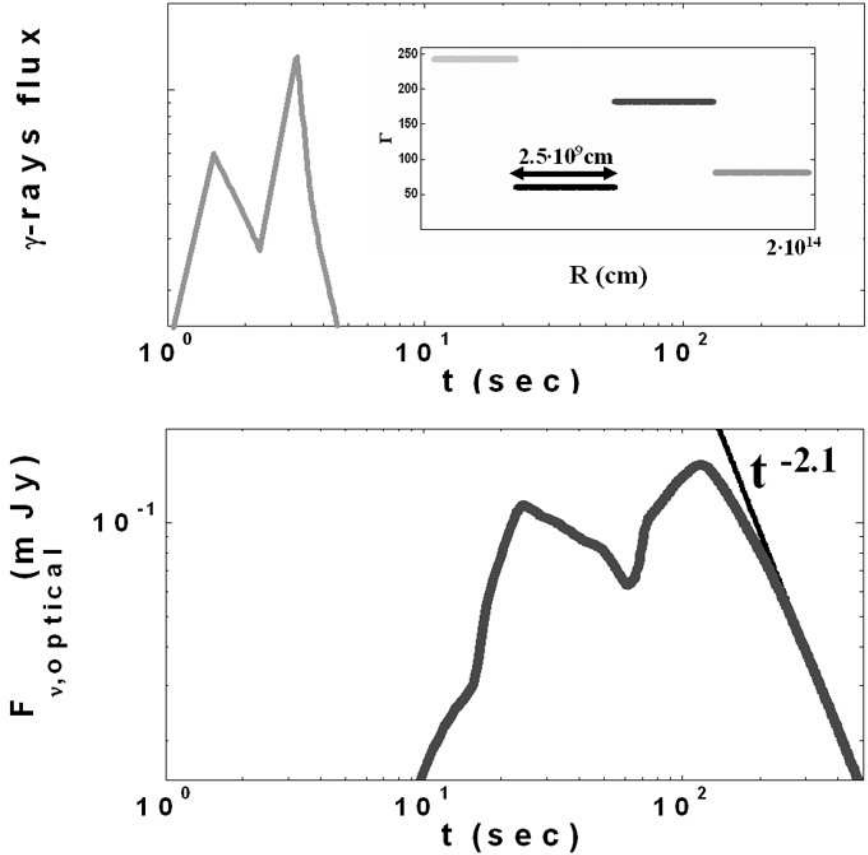}     
\caption{ {\bf Top}: An illustration of the prompt $\gamma$-ray
emission (the flux is in arbitrary units) resulting from the
collisions of four shells. The initial lorentz factor and width of
these four shells is depicted in the inset. {\bf Bottom}: The
optical light curve of the reverse shock as simulated numerically.
The initial hydrodynamic profile of the simulation appears in the
top panel inset. The total energy (isotropic equivalent) of the
ejecta is $8 \cdot 10^{51}$ erg and the external density is a
constant $10 \;\rm cm^{-3}$. The energy equipartition parameters of
the electrons and the magnetic field $\epsilon_e=0.5$ and
$\epsilon_B=0.01$ respectively. The electrons spectral index is
$p=2.5$ and the burst is at a redshift $z=1$. Note that the time
axis is logarithmic.}
\end{figure}

In GRBs the reverse shock is mildly relativistic (\cite{SP95}). This
implies that during one crossing the shock dissipates almost all the
bulk motion energy of the outflow to internal energy. After a single
crossing of the outflow a rarefaction wave is reflected and only the
forward shock remains, forming the shocked external medium into a
Blandford-Mckee self-similar solution. The original ejecta expands
and cools at the tail of the shocked external medium. The reverse
shock emission can be divided to two distinctive phases, the RS
crossing phase and the expanding-cooling phase. Observationally, the
two are separated by the peak of the optical flash. The evolution
before the time of the optical peak, $t_0$, is highly sensitive to
the initial profile of the ejecta, namely to the density and the
Lorentz factor of the ejecta at the beginning of the RS crossing.
Thus the light curve before and at $t_0$ can be used as a diagnostic
tool of these properties. On the other hand, the RS that crosses the
shell erases, to a large extend, the initial shell profile.
Moreover, the evolution during the expanding and cooling phase
depends only weakly on this profile (\cite{KS00}). Therefore, the
behavior after $t_0$ depends weakly on the initial conditions and as
such it provides a very unique signature of an RS emission.
Swift\footnote{http://swift.gsfc.nasa.gov/docs/swift/swiftsc.html}
is expected to provide detailed optical observation of the early
afterglow. In some cases the observation is expected to start before
$t_0$ and thus open a unique window into the properties of the
ejecta (in case that the ejecta is baryonic). Moreover, here we show
that if the prompt emission results from internal shock then the
light curve of the early optical afterglow at $t<t_0$ is expected to
be correlated with the prompt $\gamma$-ray emission (see fig. 1).

\section{The light curve at $t>t_0$ - a test of the reverse shock}
After the reverse shock crosses the ejecta, the shocked plasma cools
and expand at the tail of the shocked external medium. Kobayashi and
Sari (\cite{KS00}) explored this phase numerically and they have
shown that the evolution of the hydrodynamical parameters in the
cooling ejecta depends weekly on the strength of the reverse shock
(that in turn depends on the initial conditions of the ejecta). As a
result the light curve during this phase depends on the initial
conditions only by the relative values of the three reverse shock
break frequencies at $t_0$ ($\nu_a^r$, the self-absorbtion
frequency, $\nu_m^r$, the synchrotron frequency and $\nu_c^r$ the
cooling frequency). over a wide range of the parameter space
(assuming an ISM\footnote{If the external medium is a typical wind
from a massive star than the density is much higher and as a result
$\nu_c^r(t_0)<\nu_{opt}$ \cite{CL00}. In this case the light curve
evolution at $t>t_0$ is different than the evolution in an ISM.})
these frequencies satisfy
$\nu_{radio}<\nu_m^r(t_0)<\nu_a^r(t_0)<\nu_{opt}<\nu_c^r(t_0)$,
where $\nu_{radio}$ and $\nu_{opt}$ are the observed radio and
optical frequencies respectively (\cite{NP04}). In this case the
optical light curve decays at $t>t_0$ as $\sim t^{-2}$
(\cite{SP99}). In contrast to the optical emission, the radio
continues to rise at $t>t_0$ and it peaks at a later time, $t_*$,
when $\nu_{radio}=\nu_a^r$. Over a wide range of initial conditions
(when the shock is not ultra relativistic) $\nu_{radio}<\nu_m^r(t_0)
< \nu_a^r(t_0) \approx 10^{12-13}$Hz. In this case the radio
emission is expected to rise first as $\sim t^{0.5}$ until
$\nu_{radio}=\nu_m^r$ and then as $\sim t^{1.25}$ until $t_*$. At
$t>t_*$ it is expected to decay, similarly to the optical emission,
as $\sim t^{-2}$. This behavior implies that:
\begin{equation}\label{EQ radioTest}
\frac{F_*}{F_0}\left(\frac{t_*}{t_0}\right)^{\frac{p-1}{2}+1.3}=C\left(\frac{\nu_{opt}}{\nu_{radio}}\right)^\frac{p-1}{2}
\sim 1000,
\end{equation}
where $F_0$[$F_*$] is the peak optical [radio] flux (at
$t_0$[$t_*$]). $C$ is a factor of the order unity that arises due to
uncertainty in the hydrodynamics (\cite{KS00}). This entire
evolution provides several independent tests to verify that we
observe a reverse shock emission. It provides also an independent
measurement of $\nu_a^r(t_0)$:
\begin{equation}\label{EQ nuat0}
\nu_a^r(t_0)\approx \frac{t_*}{t_0} \nu_{radio}.
\end{equation}
Detailed radio observations at $t<t_*$ that identify the break
during the rising phase, when $\nu_m^r=\nu_{radio}$ would enable a
determination of $\nu_m^r(t_0)$ as well.

\section{Internal shocks and the reverse shock light curve at $t<t_0$}
The reverse shock emission during the time that the shock crosses
the ejecta depends strongly on the strength of the shock, and in
particular on the value of $n_0\Gamma^2/n_{ejecta}$ where $n_0$ is
the density of the ISM and $\Gamma$ [$n_{ejecta}$] is the Lorentz
factor [density] of the ejecta. The density profile of the ejecta
before the internal shocks is not smoothed by the shocks, on the
contrary, a density contact discontinuity is produced whenever an
internal shock occurs within the ejecta. Therefore at the end of the
internal shocks the density profile of the ejecta reflects the
initial shells that produced the internal shocks. Since the {\it
instantaneous} emission of the RS depends strongly on the density of
the ejecta it would vary strongly whenever the RS crosses a density
discontinuity that remains from the internal shocks. As a result of
the angular smoothing (\cite{FMN95,SP97}) the {\it observed} RS
emission is an average of the {\it instantaneous} emission over a
range of radii. This smoothing limits the variability time scale of
the RS to be of the order of the observed time since the beginning
of the prompt emission. Therefore, if the prompt $\gamma$-ray
emission results from internal shocks, the RS optical light curve at
$t<t_0$ is expected to resemble a smooth version of the prompt
$\gamma$-ray emission.

In order to demonstrate this relation between the prompt
$\gamma$-ray emission and the reverse shock light curve at $t<t_0$
we have carried out a one dimensional hydrodynamic numerical
simulation\footnote{The hydrodynamics simulations were done using a
one dimensional Lagrangian code that was provided to us generously
by Re'em Sari and Shiho Kobayashi (\cite{KS00}).}, starting before
the internal shocks and ending long after the reverse shock crossed
the ejecta. The initial conditions are four shells with Lorentz
factors that vary between 60 and 240 with constant energy (see the
inset in the top panel of fig. 1). The shells collide at a radius of
$\sim 2 \times 10^{14}$ cm and produce a prompt $\gamma$-ray
emission composed of two pulses with durations of approximately a
second each. An illustration of the prompt $\gamma$-ray emission is
depicted at the upper panel of fig. 1. Having the hydrodynamic
evolution of the ejecta and the external medium we used a detailed
synchrotron radiation code\footnote{The synchrotron radiation code
is described in \cite{NG05}} to calculate the RS emission. The
resulting optical light curve is depicted in the bottom panel of fig
1. The similarity between the two light curves is clear. Note also
that in this specific case the optical reverse shock emission peaks
long after the prompt $\gamma$-ray emission ends. This delay is
expected when the ejecta is thin. In this case the ejecta expands
linearly with the radius between the internal shocks radius ($\sim
10^{14}$ cm) and the external shock radius ($\sim 10^{16}$ cm).
Figure 1 shows also that after the reverse shock crosses the ejecta,
$t>t_0$, the optical light curve is insensitive to the complex
initial profile of the ejecta and it decays as $t^{-2.1}$.

We thank R. Sari and S. Kobayashi for providing us with their
relativistic hydrodynamics code and J. Granot, R. Mochkovitch, F.
Daigne  and  E. Rossi for helpful discussions.

\end{document}